%% file: main.tex
\documentclass[sigconf]{acmart}

\usepackage{color}
\usepackage{float}
\usepackage{subcaption}
\usepackage{xspace} 
\usepackage{multirow}
\usepackage{array}
\usepackage{enumitem}
\usepackage{soul}

\definecolor{burntorange}{rgb}{0.8, 0.33, 0.0}
\definecolor{byzantium}{rgb}{0.44, 0.16, 0.39}
\definecolor{byzantine}{rgb}{0.74, 0.2, 0.64}
\definecolor{lightlightgray}{rgb}{0.94, 0.94, 0.94}
\definecolor{lightblue}{rgb}{0.91, 0.95, 0.99}

\AtBeginDocument{%
  \providecommand\BibTeX{{%
    \normalfont B\kern-0.5em{\scshape i\kern-0.25em b}\kern-0.8em\TeX}}}

\copyrightyear{2024}
\acmYear{2024}
\setcopyright{rightsretained}
\acmConference[CHI EA '24]{Extended Abstracts of the CHI Conference on Human Factors in Computing Systems}{May 11--16, 2024}{Honolulu, HI, USA}
\acmBooktitle{Extended Abstracts of the CHI Conference on Human Factors in Computing Systems (CHI EA '24), May 11--16, 2024, Honolulu, HI, USA}
\acmDOI{10.1145/3613905.3650756}
\acmISBN{979-8-4007-0331-7/24/05}

\newcommand{\systemname}{\textsc{ConstraintMaker}\xspace} % SYSTEMNAME

\newcommand{\userquote}[1]{``\textit{#1}''}

\newcommand{\hlc}[2][yellow]{{%
    \colorlet{foo}{#1}%
    \sethlcolor{foo}\hl{#2}}%
}

\newcommand{\constraintprimitive}[1]{\hlc[lightlightgray]{\small\texttt{~#1~}}}

\begin{document}


% v3
\title{``We Need Structured Output'': Towards User-centered Constraints on Large Language Model Output}

% v2
% \title{``Make Sure to Output in JSON'': Towards a User-centered Perspective on Constraining LLM Output}

% v1
% \title{Toward a User-centered Perspective on Constraining LLM Output: Use Cases, Benefits, and Preferences}

\author{Michael Xieyang Liu}
\affiliation{%
  \institution{Google Research}
  \city{Pittsburgh, PA}
  \country{USA}
  }
\email{lxieyang@google.com}

\author{Frederick Liu}
\affiliation{%
  \institution{Google Research}
  \city{Seattle, Washington}
  \country{USA}
  }
\email{frederickliu@google.com}

\author{Alexander J. Fiannaca}
\affiliation{%
  \institution{Google Research}
  \city{Seattle, Washington}
  \country{USA}
  }
\email{afiannaca@google.com}

\author{Terry Koo}
\affiliation{%
  \institution{Google}
  \city{Indiana}
  \country{USA}
  }
\email{terrykoo@google.com}

\author{Lucas Dixon}
\affiliation{%
  \institution{Google Research}
  \city{Paris}
  \country{France}
  }
\email{ldixon@google.com}

\author{Michael Terry}
\affiliation{%
  \institution{Google Research}
  \city{Cambridge, Massachusetts}
  \country{USA}
  }
\email{michaelterry@google.com}

\author{Carrie J. Cai}
\affiliation{%
  \institution{Google Research}
  \city{Mountain View, California}
  \country{USA}
  }
\email{cjcai@google.com}

\renewcommand{\shortauthors}{Liu et al.}

\begin{abstract}
Large language models can produce creative and diverse responses. However, to integrate them into current developer workflows, it is essential to constrain their outputs to follow specific formats or standards. 
In this work, we surveyed 51 experienced industry professionals to understand the range of scenarios and motivations driving the need for output constraints from a user-centered perspective. 
We identified 134 concrete use cases for constraints at two levels: low-level, which ensures the output adhere to a structured format and an appropriate length, and high-level, which requires the output to follow semantic and stylistic guidelines without hallucination.
Critically, applying output constraints could not only streamline the currently repetitive process of developing, testing, and integrating LLM prompts for developers, but also enhance the user experience of LLM-powered features and applications.
We conclude with a discussion on user preferences and needs towards articulating intended constraints for LLMs, alongside an initial design for a constraint prototyping tool.

\end{abstract}





%  \includegraphics[width=\textwidth]{assets/images/fig-list-view-crop.pdf}
%  \caption{\systemname's list view UI. As the developer browses a web page (b), \systemname attempts to automatically collect options and criteria from the page, and display them in the option (c) and criteria pane (d) in the sidebar (a). In addition, \systemname leverages natural language processing to automatically group similar criteria together (e) as well as behavioral signals such as mouse movement and dwell time to automatically figure out the order of importance of the criteria. Users can use the ``See more'' and ``See less'' buttons (g) to adjust how many criteria are to be displayed at once. The sidebar can be toggled in and out by clicking the browser extension icon (h).}
%  \Description{Teaser Figure}
%  \label{fig:teaser-image}

\maketitle

\input{sections/sec-010-intro}
\input{sections/sec-020-methods}

% \input{sections/sec-030-demographics}
\input{sections/sec-031-use-cases}
\input{sections/sec-032-importance}
\input{sections/sec-033-gui-nl}

\input{sections/sec-040-tool}

% \input{sections/sec-050-discussion}

\input{sections/sec-060-conclusion}

% \begin{acks}

% \end{acks}

\bibliographystyle{ACM-Reference-Format}
\bibliography{references}

\clearpage
\newpage
\onecolumn
\appendix

\input{sections/apx-methodology}

\end{document}

%% file: sections/sec-010-intro.tex
\section{Introduction}

Over the past few years, we have witnessed the extraordinary capability of Large Language Models (LLMs) to generate responses that are not only creative and diverse but also highly adaptable to various user needs \cite{ouyang_training_2022,brown_language_2020,chowdhery_palm_2022,petridis_promptinfuser_2023,liu_what_2023,liu_selenite_2023,petridis_constitutionmaker_2023,liu_tool_2023}. 
% This accessibility has empowered scientists, developers, entrepreneurs, and enthusiasts to easily and rapidly prototype and build AI functionalities \cite{jiang_promptmaker_2022,jiang_genline_2021,petridis_constitutionmaker_2023,park_generative_2023} without extensive requirements in machine learning (ML) or natural language processing (NLP) expertise \cite{wu_promptchainer_2022,wu_ai_2022,terry_ai_2023,jiang_promptmaker_2022}. 
For example, researchers can prompt ChatGPT \cite{openai_chatgpt_2023} to condense long articles into concise summaries for fast digestion; while video game developers can generate  detailed character profiles with rich personality traits, backstories, and unique abilities on demand, simply by dynamically prompting an LLM with the game context and players' preferences. % can cut these examples if they are not essential

As much as end-users appreciate the unbounded creativity of LLMs, recent field studies examining the development of LLM-powered applications have repeatedly demonstrated the necessity to \emph{impose constraints} on LLM outputs \cite{parnin_building_2023,fan_large_2023}. For instance, a user might require a summary of an article to be ``strictly less than 20 words'' to meet length constraints, or a generated video game character profile to be ``a valid JSON that can be parsed by Python'' for a development pipeline.\looseness=-1

However, as evidenced by many recent NLP benchmarks and evaluations \cite{sun_evaluating_2023,zhou_instruction-following_2023,zeng_evaluating_2023,kahng_llm_2024}, current state-of-the-art LLMs still lack the ability to \emph{guarantee} that the generated output will invariably conform to user-defined constraints in the prompt (sometimes referred to as \emph{controllability}). Although researchers have proposed various methods to improve \textit{controllability}, such as supervised fine-tuning with specialized datasets \cite{sun_aesop_2021} or controlled decoding strategies \cite{xu_look-back_2023,beurer-kellner_prompting_2023,mudgal_controlled_2023,noauthor_guidance-aiguidance_2023}, they tend to only focus on addressing a narrow range of constraints without taking into consideration the diverse usage scenarios and rationale that real-world developers and end-users encounter when prototyping and building practical LLM-powered functionalities and applications \cite{lu_neurologic_2022,hokamp_lexically_2017,hu_improved_2019,jiang_promptmaker_2022}.

In this work, we took the first step to systematically investigate the scenarios and motivations for applying output constraints from a \emph{user-centered perspective}. Specifically, we sought to understand: 

\begin{itemize}[leftmargin=.25in]
    \item \textbf{RQ1}: What real-world use cases would necessitate or benefit from being able to constrain LLM outputs?
    
    \item \textbf{RQ2}: What are the benefits and impacts of being able to apply constraints to LLM outputs?
    % The importance of having constrained output and subsequent impact on their workflows;
    
    \item \textbf{RQ3}: How would users like to articulate their intended constraints to LLMs?
\end{itemize}

We investigated these research questions by distributing a survey to a broad population of industry professionals (software engineers, researchers, designers, project managers, etc.) who have experience building LLM-powered applications. 
Our analysis identified six primary categories of output constraints that users desire, each supported by detailed usage scenarios and illustrative examples, summarized in Table \ref{tab:use-case}. In a nutshell, users not only need \emph{low-level constraints}, which mandate the output to conform to a structured format and an appropriate length, but also desire \emph{high-level constraints}, which involve semantic and stylistic guidelines that users would like the model output to adhere to without hallucinating. 
Notably, developers often have to write complex code to handle ill-formed LLM outputs, a chore that could be simplified or eliminated if LLMs could strictly follow output constraints. In addition, the ability to apply constraints could help ease the integration of LLMs with existing pipelines, meet UI and product specifications, and enhance user trust and experience with LLM-powered features. 
Moreover, we discovered that describing constraints in natural language (NL) within prompts is not always the preferred method of control for LLM users. Instead, they seek alternative mechanisms, such as using graphical user interfaces, or GUIs, to define and test constraints, which could offer greater flexibility and a heightened sense of assurance that the constraints will be strictly followed. 

Informed by these results, we present an early design of \systemname, a prototype tool that enables LLM users to experiment, test, and apply constraints on the format of LLM outputs (see Figure \ref{fig:tool} for more details), along with feedback and insights from preliminary user tests. Overall, this paper contributes:

% \vspace{-3mm}
\begin{itemize}[leftmargin=.25in]
    \item a comprehensive taxonomy summarizing both low-level and high-level output constraints desired by LLM users (Table \ref{tab:use-case}), derived from 134 real-world use cases reported by our survey respondents (RQ1),
    
    \item an overview of both developer and user-facing benefits of being able to impose constraints on LLM outputs (RQ2),
    
    \item an exploration of LLM users' preferences for expressing constraints, whether via GUIs or natural language (RQ3),
    
    \item a initial design of the tool \systemname, which enables users to visually prototype LLM output constraints, accompanied by a discussion of preliminary user feedback.
\end{itemize}

% Overall, this paper contributes: 1) insights from a survey of LLM practitioners that reveal a diverse range of use cases, motivations, and preferences for implementing constraints on LLM outputs; 2) a preliminary design for a tool that allows LLM practitioners to prototype and evaluate constraints, supplemented by initial user feedback.

% \temp{Based on the survey, we further created a GUI + controlling mechanism to exercise control, bridging the gap.}

% Recent work in ML and NLP has proposed various methods to improve the \textit{controllability} of language models, such as supervised finetuning with carefully-curated datasets \needcite or constrained decoding strategies \needcite. In the meantime, researchers have handcrafted various benchmarks to evaluate said controllability, such as Numerical Planning Benchmark \cite{sun_evaluating_2023} and IFEval \cite{zhou_instruction-following_2023}. 

% \temp{
% to combat, few shot (not coverage), and (see hci research proposal) --- a gap between user's intent for control and the ability for the model to interpret and follow those controls during generation.
% }

% \temp{
% a lack of understanding of what possible constraints that end-users want, how would that impact their workflows, and how they prefer to express those constraints.
% }

%% file: sections/sec-020-methods.tex
\section{Survey with Industry Professionals}\label{sec:methods}

\paragraph{Methodology}
To get a broad range of insights from people who have experience with prompting and building LLM-powered applications, we deployed an online survey to users of an internal prompt-based prototyping platform (similar to the OpenAI API Playground \cite{openai_playground_2023} and Google AI studio \cite{google_google_2023}) at a large technology company for two weeks during Fall 2023. We chose this platform because it was explicitly designed to lower the barriers to entry into LLM prompting and encourage a broader population (beyond machine learning professionals) to prototype and develop LLM-powered applications. 
We publicized the survey through the platform's user mailing list. We ran the survey for two weeks during Fall 2023. Participants were rewarded \$10 USD for their participation. The survey was approved by our organization's IRB.
% Participants were required to be at least 18 years old and fluent in English.

% \subsection{The Survey Instrument}

\paragraph{Instrument}
Our survey started with questions concerning participants' background and technical proficiency, such as their job roles and their level of experience in designing and engineering LLM prompts. 
The survey subsequently investigated RQ1 and RQ2 by asking participants to report \emph{three real-world use cases} in which they believe the implementation of constraints to LLM outputs is necessary or advantageous. For each use case, they were encouraged to detail the specific scenario where they would like to apply constraints, the type of constraint that they would prefer to implement, the degree of precision required in adhering to the constraint, and the importance of this constraint to their workflow.
Finally, the survey investigated RQ3 by asking participants to reflect on scenarios where they would prefer \emph{expressing constraints via a GUI} (sliders, buttons, etc.) over natural language (in prompts, etc.) and vice versa, as well as any alternative ways they would prefer to articulate constraints to LLMs. The GUI alternative draws inspiration from tools like the OpenAI Playground that allow users to adjust settings like temperature and top-k through buttons, toggles, and sliders.
Detailed survey questions are documented in section \ref{sec:apx-survey-instrument} of the Appendix.

\paragraph{Results}
51 individuals responded to our survey. Over half of the respondents were software engineers (58.8\%) across various product teams; others held a variety of roles like consultant \& specialist (9.8\%), analyst (7.8\%), researcher (5.9\%), UX engineer (5.9\%), designer (3.9\%), data scientist (3.9\%), product manager (2.0\%), and customer relationship manager (2.0\%).
All respondents had experience with prompt design and engineering, with the majority reported having extensive experience (62.7\%).
The targeted audience and use cases of their prompts were split approximately evenly among consumers and end-users (33.3\%), downstream development teams (31.4\%), or both (29.4\%), with some created specifically for exploratory research \& analysis (5.9\%).
Together, respondents contributed 134 unique use cases of output constraints. To analyze the contents of the open-ended responses, the first author read through all responses and used inductive analysis \cite{strauss_basics_1990} to generate and refine themes for each research question with frequent discussions with the research team. We present the resulting themes for each research question in sections \ref{sec:use-cases}-\ref{sec:gui-or-nl}.

\input{use-cases-table}

\paragraph{Limitations}
Note that our findings largely capture the views of industry professionals, and may not encompass those of casual users who engage with LLMs conversationally \cite{openai_chatgpt_2023}. Additionally, as our respondent sample is limited to a single corporation, the results described in the following sections may not be representative of the entire industry. Furthermore, our frequent use of open-ended questions might have negatively impacted the response rate. However, the saturation of novel findings and insights towards the end of the survey deployment suggests that we have successfully captured a comprehensive range of perspectives.

% original:
% \temp{
% The survey was primarily focused on industry professionals experienced in prompting and building LLMs. Consequently, our respondents may not represent the entire spectrum of LLM users. That said, we postulate that industry professionals, including software engineers and UX designers, may express a much stronger desire for output constraints on LLMs, which would facilitate the integration of LLMs into existing systems and workflows. Output constraint might likely to be of lesser concern for non-professional or amateur users who interact with LLMs more casually, e.g., through chat interfaces. The exploration of this particular user group, however, is beyond the scope of our current study.

% In addition, compared to other large-scale surveys \cite{cai_software_2019}, we only garnered a relatively small number of individual responses, which might limit the statistical power and generalizability of the findings.
% A potential obstacle could be attributed to our survey design --- we largely relied on open-ended questions, which, while helpful for capturing a variety of perspectives, placed a heavy cognitive and physical load on the respondents to fill out.
% However, one encouraging observation was that towards the end of the deployment, the influx of novel and unique findings began to taper off, suggesting we had reached a point of saturation and thus a reasonable stopping point for our survey.
% }

%% file: use-cases-table.tex
\newcommand{\dottedQuote}[1]{\textit{``#1''}}
\def\arraystretch{1.3}
\begin{table*}[t]
\centering
% \vspace{-2mm}
\resizebox{1\textwidth}{!}{%
\begin{tabular}{
p{20mm}
p{34mm}
p{6mm}
p{111mm}
p{12mm}
}
\toprule

\multicolumn{2}{l}{{\textsc{Category}}} & 
\textsc{\%} & 
{\textsc{Representative Examples}} %\\\midrule
& 
{\textsc{Precision}} \\\midrule

\multicolumn{5}{l}{{\emph{Low-level constraints}}} \\\midrule
% 
% 
% 
% Structured Output
% \multicolumn{2}{l}{\textbf{Structured Output}} & 
% \multicolumn{3}{l}{\textbf{\temp{xx\%}}} \\\midrule
\multirow{2}{\linewidth}{\vspace{-17mm}\parbox{\linewidth}{\textbf{Structured \newline Output}}}
& 
Following standardized or custom format or template (e.g., markdown, HTML, DSL, bulleted list, etc.) & 
26.1\% & 
\dottedQuote{Summarizing meeting notes into markdown format} \newline \dottedQuote{... the chatbot should quote dialogues, use special marks for scene description, etc.} \newline \dottedQuote{I want the output to be in a specific format for a list of characteristics [of a movie] to then easily parse and train on.}
&
Exact \\

& 
Ensuring valid JSON object (with custom schema) & 
16.4\% & 
\dottedQuote{... use the JSON output of the LLM to make an http request with that output as a payload.} \newline \dottedQuote{I want to have the output [of the quiz] to be like a json with keys \texttt{\{"question": "...", "correct\_answer": "...", "incorrect\_answers": [...]\}}} %\\\midrule
&
Exact \\
\midrule

% 
% 
% 
% Multiple Choice
% \multicolumn{2}{l}{\textbf{Multiple Choice}} & 
% \multicolumn{3}{l}{\textbf{\temp{xx\%}}} \\\midrule
\textbf{Multiple \newline Choice} & 
Selecting from a predefined list of options & 
23.9\% & 
\dottedQuote{Classifying student answers as right / wrong / uncertain...)} \newline \dottedQuote{While doing sentimental analysis, [...] restrict my output to few fixed set of classes like Positive, Negative, Neutral, Strongly Positive, etc.} %\\\midrule
&
Exact \\\midrule

% 
% 
% 
% Length Constraints
% \multicolumn{2}{l}{\textbf{Length Constraints}} & 
% \multicolumn{3}{l}{\textbf{\temp{xx\%}}} \\\midrule
\textbf{Length \newline Constraints}\newline & 
Specifying the targeted length (e.g., \# of characters / words, \# of items in a list) & 
16.4\% & 
\dottedQuote{... Make each summary bullet LESS THAN 40 words. If you generate a bullet point that is longer than 40 words, summarize and return a summary that is 40 words or less.} \newline \dottedQuote{I want to limit the characters in the output to 100 so it is a valid YouTube Shorts title.} %\\\midrule
&
Approx. \\\midrule

\multicolumn{4}{l}{{\emph{High-level constraints}}} \\\midrule
% 
% 
% 
% Semantic Constraints
% \multicolumn{2}{l}{\textbf{Semantic Constraints}} & 
% \multicolumn{3}{l}{\textbf{\temp{xx\%}}} \\\midrule
\multirow{4}{\linewidth}{\vspace{-38mm}\parbox{\linewidth}{\textbf{Semantic \newline Constraints}}} &
Excluding specific terms, items, or actions & 
8.2\% & 
\dottedQuote{Exclude PII (Personally Identifiable Information)  and even some specific information...} \newline \dottedQuote{If asking for html, do not include the standard html boilerplate (doctype, meta charset, etc.) and instead only provide the meaningful, relevant, unique code.} %\\
&
Exact \\

& 
Including or echoing \newline specific terms or content & 
2.2\% &
\dottedQuote{... I want [the email] to include about thanking my manager and also talk about the location he is based on to help him feel relatable.} \newline \dottedQuote{We want LLM to repeat input with some control tokens to indicate the mentions. e.g. input: `Obama was born in 1961.',... , we want output to be `<<Obama>> was born in 1961.'} %\\
& 
Exact \\
 
& 
Covering or staying on a \newline certain topic or domain & 
2.2\% & 
\dottedQuote{[The output of] a query about `fall jackets' should be confined to clothing.} \newline \dottedQuote{For ex. In India we have Jio and Airtel as 2 main telecom service provider. While building chat bot for Airtel, I would want the model to only respond [with] Airtel related topics.} %\\
& 
Exact \\

& 
Following certain (code) \newline grammar / dialect / context & 
4.5\% & 
\dottedQuote{While generating SQL,... restrict the output to a particular dialect and use the table / database name mentioned in the prompt.} \newline \dottedQuote{... implement[ing] a voice assistant that calls specific methods with relevant arguments,... the output needs to be valid syntax and only call the methods specified in the context} %\\\midrule
& 
Exact \\\midrule

% 
% 
% 
% Stylistic Constraints
% \multicolumn{2}{l}{\textbf{Stylistic Constraints}} & 
% \multicolumn{3}{l}{\textbf{\temp{xx\%}}} \\\midrule
\textbf{Stylistic \newline Constraints} & 
Following certain style, tone, or persona & 
6.7\% & 
\dottedQuote{... it is important that the [news] summary follow a style guide, ... for example, preference for active voice over passive voice.} \newline \dottedQuote{Use straightforward language and avoid complex technical jargon...} %\\\midrule 
&
Approx. \\\midrule

% 
% 
% 
% Prevent Hallucination
% \multicolumn{2}{l}{\textbf{Preventing Hallucination}} & 
% \multicolumn{3}{l}{\textbf{\temp{xx\%}}} \\\midrule
\multirow{2}{\linewidth}{\vspace{-12mm}\parbox{\linewidth}{\textbf{Preventing Hallucination}}} & 
Staying grounded and \newline truthful & 
8.2\% & 
\dottedQuote{... we do not want [a summary of the doc] to include opinions or beliefs but only real facts.} \newline \dottedQuote{If the LLM can't find a paper or peer-reviewed study, do not provide a hallucinated output.} %\\
&
Exact \\

& Adhering to instructions (without improvising \newline unrequested actions) & 
4.5\% & 
\dottedQuote{For `please annotate this method with debug statements', I'd like the output to ONLY include changes that add print statements... No other changes in syntax should be made.
} \newline \dottedQuote{LLMs usually ends up including an advice associated to the summarised topic, advice we need to avoid so they are not part of the doc.}  %\\\bottomrule
&
Exact \\\bottomrule

% % 
% % 
% % 
% % Others
% % \multicolumn{2}{l}{\textbf{Others}} & 
% % \multicolumn{3}{l}{\textbf{\temp{xx\%}}} \\\midrule
% \multirow{2}{\linewidth}{\textbf{Others}\newline\temp{xx\%}} & 
% Stop upon satisfaction & 
% \temp{xx\%} & 
% \dottedQuote{[a model that predicts code] should stop when it sees a certain word (or one of [the] words).} \\
% % &
% % Exact \\

% & Multi-modal output order & 
% \temp{xx\%} & 
% \dottedQuote{I want images and text outputted in a pre-determined order.} \\\bottomrule
% % &
% % Exact \\\bottomrule
\end{tabular}
}
\vspace{0mm}
\caption{Taxonomy of the six primary categories of use cases of output constraints, derived from the 134 use cases that respondents submitted (RQ1). Totals add up to more than 100\% since we placed some use cases into more than one category. 
The final column indicates whether the output is expected to match the constraint \textit{exactly} or \textit{approximately}, as agreed upon by the majority of respondents.
}
\label{tab:use-case}
\vspace{-3mm}
\end{table*}

%% file: sections/sec-031-use-cases.tex
\section{RQ1: Real-world Use Cases that Necessitate Output Constraints}\label{sec:use-cases}

Table \ref{tab:use-case} presents a taxonomy of six primary categories of use cases that require output constraints, each with representative real-world examples and quotes submitted by our respondents. 

These can be further divided into \emph{low-level} and \emph{high-level} constraints --- low-level constraints ensure that model outputs adhere to a specific \textbf{structure} (e.g., JSON or markdown), instruct the model to perform pure \textbf{multiple choices} (e.g., sentiment classification), or dictate the \textbf{length} of the outputs; whereas high-level constraints enforce model outputs to respect \textbf{semantic} (e.g., must include or avoid specific terms or actions) or \textbf{stylistic} (e.g., follow certain style or tone) guidelines, while \textbf{preventing hallucination}. 

Below, we discuss a number of interesting insights that emerged from our analysis of the use cases:

\input{importance-table}

\begin{itemize}[leftmargin=.15in,itemsep=0.05in]
    \item 
    \textbf{Going beyond valid JSON.} 
    % A substantial portion of respondents (\temp{xx\%}) mentioned that they would like LLM outputs to be \emph{valid JSON objects} that can be seamlessly \userquote{parsed and integrated into downstream applications.} 
    Note that recent advancements in instruction-tuning techniques have substantially improved the the chances of generating a valid JSON object upon user request \cite{openai_json_2023,ouyang_training_2022}. Nonetheless, our survey respondents believed that this was not enough and \textbf{desired to have more precise control over the JSON schema (i.e., key/value pairs)}. One respondent stated their expectation as follows: \userquote{I expect the quiz [that the LLM makes given a few passages provided below] to have 1 correct answer and 3 incorrect ones. I want to have the output to be like a json with keys \texttt{\{"question": "...", "correct\_answer": "...", "incorrect\_answers": [...]\}}.} 
    It is also worth mentioning that some respondents
    found that \userquote{few-shot prompts} --- demonstrating the desired key/value pairs with several examples --- tend to work \userquote{fairly well}. However, they concurred that having a formal guarantee of JSON schema would be greatly appreciated  (see section \ref{sec:importance-increase-prompt-dev-efficiency} for their detailed rationales).

    \item 
    \textbf{Giving an answer without extra conversational prose.} 
    When asking an LLM to perform data classification or labeling, such as \userquote{[classifying sentiments as] Positive, Negative, Neutral, etc.,} respondents typically expect the model to \textbf{only output the classification result} (e.g. ``\texttt{Positive.}'') \textbf{without a trailing ``explanation''} (e.g., ``\texttt{Positive, since it referred to the movie as a `timeless masterpiece'...}''), as the addition of explanation could potentially confuse the downstream parsing logic. This indicates a potential misalignment between a common training objective --- where LLMs are often tailored to be conversational and provide rich details \cite{stiennon_learning_2020,bai_training_2022,liang_holistic_2023} --- and certain specialized downstream use cases where software developers need LLMs to be succinct. Such use cases necessitate output constraints that are independent of the prompt that would help adapt a general-purpose model to meet specific user requirements.\looseness=-1

    \item 
    \textbf{Conditioning the output on the input, but don't ``improvise!''} 
    One thread of high-level constraints places emphasis on directing the model to \textbf{condition its output on specific content from the input}. For example, the model's response should semantically remain \userquote{in the same ballpark} as \userquote{the user's original query} --- \userquote{[the output of] a query about `fall jackets' should be confined to clothing.} A particular instance of this is for LLMs to \emph{echo} segments of the input in their output, occasionally with slight alterations. For example, \userquote{we want LLM to repeat input with some control tokens to indicate the mentions. e.g. input: `Obama was born in 1961.',... , we want output to be `<<Obama>> was born in 1961.'} 
    Nevertheless, respondents underscored the importance of the model \textbf{not improvising beyond its input and instructions}. For example, one respondent instructed an LLM to \userquote{annotate a method with debug statement,} anticipating the output would \userquote{ONLY include changes that add print statements to the method.} However, the LLM would frequently introduce additional \userquote{changes in syntax} that were unwarranted. 
    % We further explore the importance of constraining models from such improvisation or hallucination in section \ref{sec:importance-ux}. 
    % \temp{consider including \userquote{We have a pre-set list of potential APIs that could be linked, so pre-determining the output space can help prevent the LLM from proposing unrealistic suggestions.}}
    
\end{itemize}

%% file: importance-table.tex
\def\arraystretch{1.2}
\begin{table*}[t]
\centering
% \vspace{-2mm}
\resizebox{1\textwidth}{!}{
\begin{tabular}{
p{1mm}
p{103mm}|
p{1mm}
p{68mm}
}
\toprule
\multicolumn{2}{l|}{\textsc{Developer-facing Benefits}} &
\multicolumn{2}{l}{\textsc{User-facing Benefits}} \vspace{1.2mm}\\\midrule

\multicolumn{2}{l|}{\textbf{Increasing prompt-based development efficiency (\S\ref{sec:importance-increase-prompt-dev-efficiency})}} &
\multicolumn{2}{l}{\textbf{Satisfying product and UI requirements (\S\ref{sec:importance-satisfying-ui})}} \\

~ &
Speeding up prompt design / engineering (less trial and error) \newline 
Reducing or eliminating ad-hoc parsing and plumbing logic \newline 
Saving the cost of requesting multiple candidates &
~ &
Fitting output into UI presets with size bounds \newline 
Ensuring consistency of output length and format \newline 
Complying with product and platform requirements \\\midrule

\multicolumn{2}{l|}{\textbf{Streamlining integration with downstream processes and workflows  (\S\ref{sec:importance-streamline-integration-with-downstream})}} &
\multicolumn{2}{l}{\textbf{Improving user experience and trust  (\S\ref{sec:importance-ux})}} \\

~ &
Ensuring successful code execution \newline
Improving the quality of training data synthesis \newline
Canonizing output format across models &
~ &
Eliminating safety and privacy concerns \newline
Improving user trust and confidence \newline
Increasing customer satisfaction and adoption\\\bottomrule
\end{tabular}
}
\vspace{0mm}
\caption{Respondents' perceived benefits of having the ability to apply constraints to LLM output (RQ2).}
\label{tab:importance-table}
\vspace{-5mm}
\end{table*}

%% file: sections/sec-032-importance.tex
\section{RQ2: Benefits of Applying Constraints to LLM Outputs}\label{sec:importance}

\begin{figure*}[t]
\centering
% \vspace{-2mm}
\includegraphics[width=1\linewidth]{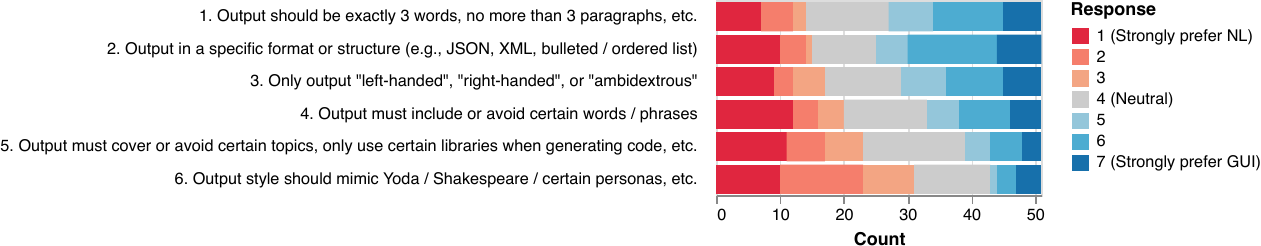}
\vspace{-6mm}
\caption{Respondents' preferences towards specifying output constraints either through natural language or GUI (RQ3). Participants were asked to score each question (left) on a 7-point Likert scale from ``1 (Strongly prefer NL)'' to ``7 (Strongly prefer GUI).''}
\label{fig:rq3-dist}
\Description{Respondents' preferences towards specifying output constraints either through natural language or GUI (RQ3). Participants were asked to score each question (left) on a 7-point Likert scale from ``1 (Strongly prefer NL)'' to ``7 (Strongly prefer GUI).'' An overarching observation is that respondents preferred using GUI to specify low-level constraints and natural language to express high-level constraints.}
% \vspace{-3mm}
\end{figure*}

Beyond the aforementioned use cases, our survey respondents reported a range of benefits that the ability of constraining LLM output could offer. These include both \emph{developer-facing} benefits, like increasing prompt-based development efficiency and streamlining integration with downstream processes and workflows, as well as \emph{user-facing} benefits, like satisfying product and UI requirements and improving user experience and trust of LLMs (Table \ref{tab:importance-table}). Here are the most salient responses:\looseness=-1

\subsection{Increasing Prompt-based Development Efficiency}\label{sec:importance-increase-prompt-dev-efficiency}

First and foremost, being able to constrain LLM outputs can significantly increase the efficiency of prompt-based engineering and development by reducing the trial and error currently needed to manage LLM unpredictability. Developers noted that the process of \userquote{defining the [output] format} alone is \userquote{time-consuming,} often requiring extensive prompt testing to identify the most effective one (consistent with what previous research has found \cite{zamfirescu-pereira_why_2023,parnin_building_2023}). Additionally, they often need to \userquote{request multiple responses} and \userquote{iterating through them until find[ing] a valid one.}
Therefore, being able to deterministically constrain the output format could not only save developers as much as \userquote{dozens of hours of work per week} spent on iterative prompt testing, but also reduce overall LLM inference costs and latency.\looseness=-1

Another common practice that respondents reported is building complex infrastructure to post-process LLM outputs, sometimes referred to as \userquote{massaging [the output] after receiving.} For example, developers oftentimes had to \userquote{chase down `free radicals' when writing error handling functions,} and felt necessary to include \userquote{custom logic} for matching and filtering, along with \userquote{further verification.} Thus, setting constraints before LLM generation may be the key to reducing such \userquote{ad-hoc plumbing code} post-generation, simplifying \userquote{maintenance,} and enhancing the overall \userquote{developer experience.} As one respondent vividly described: \userquote{it's a much nicer experience if it (formatting the output in bullets) `just works' without having to implement additional infra...}

\subsection{Integrating with Downstream Processes and Workflows}\label{sec:importance-streamline-integration-with-downstream}

% When considering LLMs as components within broader research and production pipelines, respondents emphasized the importance of output constraints in integrating LLMs with downstream process and workflows. 
Because LLMs are often used as sub-components in larger pipelines, respondents emphasized that guaranteed constraints are critical to ensuring that the output of their work is compatible with downstream processes, such as downstream modules that expect a specific format or functional code as input.
Specifically for code generation, they highlighted the necessity of constraining the output to ensure \userquote{executable} code that adheres to only \userquote{methods specified in the context} and avoids errors, such as hallucinating \userquote{unsupported operators} or  \userquote{SQL ... in a different dialect.} Note that while the ``function calling'' features in the latest LLMs \cite{openai_function_2023,google_cloud_function_2023} can ``select'' functions to call from a predefined list, users still have to implement these functions correctly by themselves.

Many studies indicate that LLMs are highly effective for creating synthetic datasets for AI training \cite{josifoski_exploiting_2023,eldan_tinystories_2023,viswanathan_prompt2model_2023}, and our survey respondents postulated that being able to impose constraints on LLMs could improve the datasets' quality and integrity. For instance, one respondent wished that model-generated movie data would \userquote{not say a movie's name when it describes its plot,} as they were going to train using this data for a \userquote{predictive model of the movie itself.} Any breach of such constraints could render the data \userquote{unusable.}

Furthermore, given the industry trend of continuously migrating to newer, more cost-effective models, respondents highlighted the importance of \userquote{canonizing} constraints across models to avoid extra prompt-engineering after migration (e.g., \userquote{if I switch model, I get the formatting immediately}). This suggests that it could be more advantageous for models to accept output constraints independent of the prompt, which should now solely contain task instructions.

\subsection{Satisfying UI and Product Requirements}\label{sec:importance-satisfying-ui}

Respondents stressed that it is essential to constrain LLM output to meet UI and product specifications, particularly when such output will be presented to end users, directly or indirectly. A common case is to incorporate LLM-generated content into UI elements that \userquote{cannot exceed certain bounds}, necessitating stringent length constraints. Content that doesn't \userquote{fit within the UI} usually gets \userquote{thrown away} all together, a concern likely to be more pronounced on mobile devices with limited screen real estate \cite{liu_wigglite_2022,chang_supporting_2016}. Maintaining the consistency of output length and format was also considered important, as \userquote{too much variability in the generated text can be overwhelming to the user and clutter the UI.} Moreover, being able to constrain length can help LLMs comply with specific platform character restrictions, like tweets capped at 280 characters or YouTube Shorts titles limited to 100 characters.

% 
% old (longer) version
% 
% Respondents suggested that the ability to constrain the output is essential for meeting established user interface (UI) and product requirements, particularly when the output from LLMs will be displayed, directly or indirectly, to end users. One scenario involves incorporating LLM-generateed content into UI components that are of a specific size, necessitating stringent length constraints: \userquote{we are trying to display a teaser to a user: In these scenarios, the text from the LLM will be displayed directly to the user in an UI that cannot exceed certain bounds. Failure to meet these bounds usually means we have to throw away all the data because we can't fit it within the UI.} This issue is likely to become more pronounced on mobile devices, which have limited screen real estate \cite{liu_wigglite_2022}. The consistency of output length and format is also essential when it comes to user-facing content, and respondents emphasized that \userquote{too much variability in the generated text can be overwhelming to the user and clutter the UI.} Moreover, with the help of constraints, LLM-generated content can more effectively comply with character limits set by certain products or platforms, such as \userquote{a tweet under 280 characters} or a valid YouTube Shorts title that \userquote{stays under 100 characters.}

\subsection{Improving User Experience, Trust, and Adoption}\label{sec:importance-ux}

Finally, respondents suggested that developing LLM-powered user experiences requires constraint mechanisms to mitigate hallucinations, foster user trust, and ultimately drive \userquote{user adoption.} One prominent aspect is to reduce safety and privacy concerns, for instance, by preventing LLMs from \userquote{repeat[ing] existing or hallucinat[ing] PII (personally identifiable information).} In addition, respondents expressed a desire to ensure user trust and confidence of LLM-powered tools and systems, arguing that, for example, \userquote{hallucinations in dates are easy to identify} and, in general, \userquote{users won't invest more time into tools that aren't accurate.}

% 
% old (longer) version
% 
% When developing user experiences powered by LLMs, respondents emphasized the importance of being able to constrain LLMs and mitigate hallucinations to enhance user experience, foster trust, and drive customer adoption. One prominent aspect is to eliminate safety and privacy concerns, for example, \userquote{it would be nice to know that the summarization will not repeat existing PII (personally identifiable information) or hallucinate some PII from the prompt.} In addition, respondents expressed a desire to ensure user trust and confidence of LLM-powered tools and systems, arguing that, for instance, \userquote{hallucinations in dates are easy to identify by the user} and \userquote{users won't invest more time into tools that aren't accurate.} Relatedly, one respondent who self-identified as a product manager wished that \userquote{it will greatly help to be accurate in all customer demos I make} in order to increase user adoption of their product.

% \userquote{replacing all occurrences of [personally identifiable information] with innocuous placeholders.}

% \temp{
% Improve user trust and confidence
% Increase customer satisfaction and adoption
% Eliminate privacy / safety concerns
% }

%% file: sections/sec-033-gui-nl.tex
\section{How to Articulate Output Constraints to LLMs}\label{sec:gui-or-nl}

\begin{figure*}[t]
\centering
% \vspace{-2mm}
\includegraphics[width=1.0\linewidth]{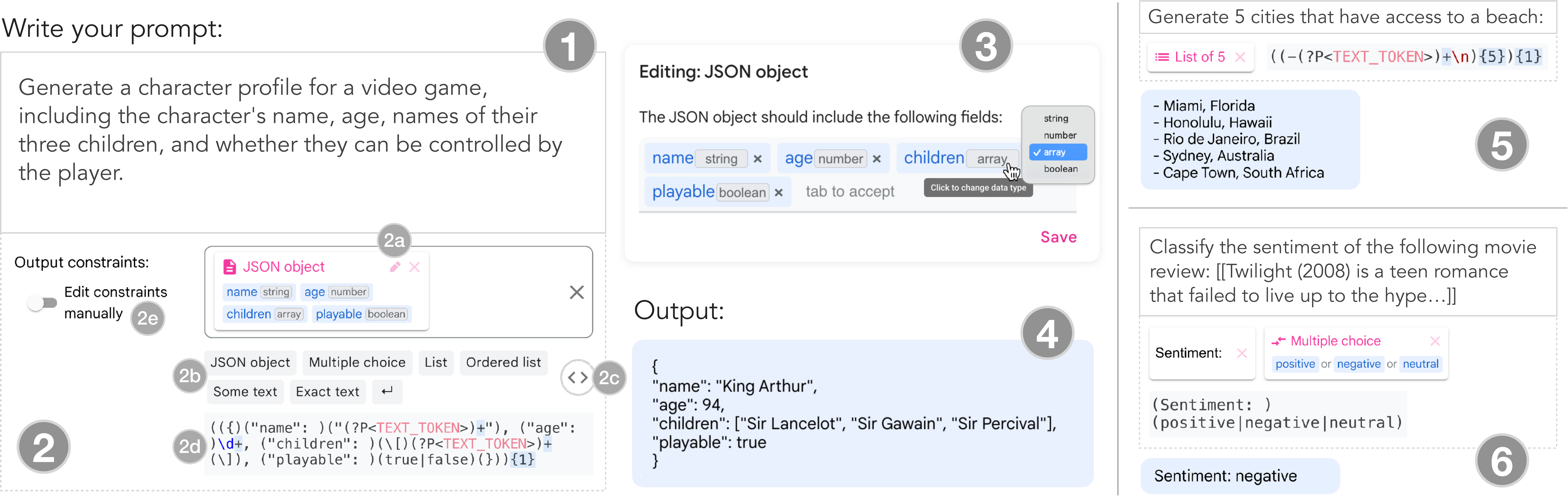}
\vspace{-7mm}
\caption{\systemname's user interfaces (1-4) \& use cases (5-6). After writing the prompt (1), users can easily specify output constraints using a graphical user interface (2 \& 3) provided by \systemname, and the resulting output (4) is guaranteed to follow the constraints. Additional details of this process is discussed in section \ref{sec:tool}.}
\label{fig:tool}
\vspace{-1mm}
\end{figure*}

% Table \ref{tab:gui-nl-table} 
Fig. \ref{fig:rq3-dist} shows distributions of respondents' preferences towards specifying output constraints either through GUI or natural language. An overarching observation is that respondents preferred \textbf{using GUI to specify low-level constraints} and \textbf{natural language to express high-level constraints}. We discuss their detailed rationale below:

\subsection{The case for GUI: A Quick, Reliable, and Flexible Way of Prototyping Constraints}

First and foremost, respondents considered GUIs particularly effective for defining \userquote{hard requirements,} providing more reliable results, and reducing ambiguity compared to natural language instructions. For example, one argued that choosing ``boolean'' as the output type via a GUI felt much more likely to be \userquote{honoured} compared to \userquote{typ[ing] that I want a Yes / No response [...] in a prompt.} Another claimed that \userquote{flagging a `JSON' button} provides a much better user experience than \userquote{typing `output as JSON' across multiple prompts.}
In addition, respondents preferred using GUI when the intended constraint is \userquote{objective} and \userquote{quantifiable}, such as \userquote{use only items x,y,z,} or \userquote{a JSON with certain fields specified.}
Moreover, respondents found GUI to be more flexible for rapid prototyping and experimentation (e.g., \userquote{when I want to play around with different numbers, moving a slider around seems easier than typing}).
Finally, for novice LLM users, the range of choices afforded by a GUI constraint can help clarify the model's capabilities and limitations, \userquote{making the model seems less like a black box.} One respondent drew from their experience working with text-to-image models to underscore this point: \userquote{by seeing ``Illustration'' as a possible output style [among others like ``Photo realistic'' or ``Cartoon''], I became aware of [the model's] capabilities.}

\subsection{The Case for NL: More Intuitive and Expressive for Complex Constraints}

Respondents found natural language easier for specifying complex constraints than GUIs, especially for extended background contexts or numerous choices that wouldn't reasonably fit into a GUI.
Natural language was also preferred for expressing vague, nuanced, or open-ended constraints, like \userquote{don't include offensive words} or \userquote{respond in a cheerful manner.} At a high level, respondents emphasized that natural language provides a more natural, familiar, and expressive way to communicate (potentially multiple) complex constraints, and, \userquote{trying to figure out how to use a GUI might be more tedious.}

% \subsection{Additional Needs}

Additionally, some respondents noted that, despite their preference for using GUIs to define constraints from time to time, they ultimately have to use natural language prompts due to API limitations. Moreover, some wished to reference \userquote{external resources} in constraints that are not feasible to directly include in prompts (e.g., \userquote{a large database / vocabulary}). These suggest that a dedicated ``output-constraints'' API field for specifying constraints could be advantageous, potentially through the use of a formal language or notation.

%% file: sections/sec-040-tool.tex
\section{The \systemname Tool}\label{sec:tool}

Informed by the survey results, we developed a web-based GUI, \systemname (Fig. \ref{fig:tool}), that enables LLM users to prototype, test, and apply constraints on the \emph{format} of LLM outputs.
With \systemname, users can specify different types of output constraints by simply selecting from the list of available \emph{constraint primitives} (Fig. \ref{fig:tool}-2b). If needed, users can click the pencil icon (Fig. \ref{fig:tool}-2a) to further edit the details of a constraint primitive, such as specifying the schema of a JSON object (Fig. \ref{fig:tool}-3). Users also have the flexibility to mix and match multiple constraint primitives together (e.g., Fig. \ref{fig:tool}-6) to form more complex constraints. Currently, based on users' needs and priorities identified by the survey, \systemname initially supports \constraintprimitive{JSON object}, \constraintprimitive{Multiple choice}, \constraintprimitive{List}, \constraintprimitive{Ordered list}, and \constraintprimitive{Some text} as primitives (Fig. \ref{fig:tool}-2b), where \constraintprimitive{Some text} asks the LLM to generate freely as it normally would.
% , and \constraintprimitive{Exact text} as primitives \ref{fig:tool}-2b, where \constraintprimitive{Some text} asks the LLM to generate freely as it normally would, and \constraintprimitive{Exact text} asks the LLM to insert user-specified text during generation.

Under the hood, we used a GPT-3.5-class LLM.
We additionally implemented a finite-state machine-based decoding technique akin to that outlined in \cite{willard_efficient_2023}, ensuring the language model outputs strictly adhere to formats defined by a specialized regular expression (henceforth, ``regex''). In fact, \systemname automatically converts a GUI-defined constraint into a regex (Fig. \ref{fig:tool}-2d), which the LLM observes during generation (Fig. \ref{fig:tool}-4).

\subsection{Iterative Design and User Feedback}
To explore the usability and usefulness of \systemname, we conducted a series of informal (around 30 minute each) user tests with five participants who self-identified as experts in prompting LLMs, as well as self-experimentation among the authors. 
% both some of the authors themselves and additional participants who self-identified as experts in prompting LLMs. 
% Sessions lasted around 30 minutes and were conducted virtually. 
We used the feedback from these sessions to iteratively refine the design of \systemname. We present some interesting findings and reflections below:

\subsubsection{\systemname enables an intuitive separation of concerns}  
With \systemname, one can now specify \userquote{the tasks they want the model to perform} \emph{separate from} \userquote{the expected format of the output,} an approach participants considered more intuitive and effective in steering LLMs to consistently achieve desired results compared to traditional prompting. 
Additionally, participants envisioned the possibility of reusing constraints across various prompts, which could reduce the effort of crafting new constraints for similar tasks and eliminate the need for prompt engineering post model migration.\looseness=-1

\subsubsection{Constraint-prototyping GUI needs to cater to both developers and non-developers} 
On the one hand, for non-technical users interested in experimenting with constraints, we noticed that the visible regex alongside the constraint primitive GUI was somewhat distracting. To address this, we added a feature that allows the regex to be toggled as hidden via the ``< >'' button (Fig. \ref{fig:tool}-2c). 
On the other hand, for more advanced users (e.g., developers), we observed a frequent need to make fine-grained adjustments to the underlying regex after creating an initial draft with the \systemname GUI (e.g., changing the ``bullet'' of a ``bulleted list'' from the default ``\texttt{$-$ [...]}'' (Fig. \ref{fig:tool}-5) to ``\texttt{* [...]}''). As a result, we enabled direct manipulation of the regex by toggling on "Edit constraints manually" (Fig. \ref{fig:tool}-2e).\looseness=-1

\subsubsection{``Inserting'' words among constraints.} 
% Participants found that sometimes it could be beneficial to ``insert'' particular words into the generation to guide the LLM and improve its output. 
For example, one participant asked in the prompt for the LLM to first write a paragraph describing a short story, followed by a list of suggestions on how to improve the story. 
In situations like this, they found that embedding specific words into the constraints, such as ``\texttt{\small Short story:} \constraintprimitive{Some text}'' followed by  ``\texttt{\small Suggestions:} \constraintprimitive{List}'', yielded better-quality results than simply using \constraintprimitive{Some text} followed by \constraintprimitive{List} alone. Therefore, we introduced the \constraintprimitive{Exact text} GUI primitive, enabling the LLM to \emph{insert} user-prescribed text into its output.

\subsubsection{Automatically inferring constraints based on prompts} 
One interesting feature request for \systemname is the ability to automatically infer constraints from user-written prompts, similar to previous intelligent prediction or auto-completion systems and tools \cite{liu_crystalline_2022,tan_visual_2023,bar-yossef_context-sensitive_2011}. For instance, for a prompt shown in Fig. \ref{fig:tool}-1, \systemname could \emph{proactively suggest} to the users if they'd like to constrain the model output to a JSON object with specific fields. 
This feature would be appealing, given the current somewhat cumbersome process of manually creating and modifying constraints from scratch. Similar to code auto-completion, participants suggested that constraint auto-completion could streamline the overall experience of defining constraints.
Additionally, automatic constraint suggestions could serve as learning opportunities for novice users to become familiar with the range of possibilities that \systemname affords, which would be particularly useful in future versions where the tool might support a wider selection of constraint primitives.
Finally, proactively suggesting constraints could promote a ``constraint mindset.'' This encourages users to always consider the output format before deploying a prompt, leading to more rigorous and controllable prompt engineering, much like conventional software development.

% However, current process of coming up with an effective constraint for a prompt can still be haphazard. Participants found that they often have to tweak the constraint and the prompt simultaneously to have them ``sync up'' 
% orchestration

% can we achieve a true separation of concern needs additional future research.

% \temp{match constraint with prompt, but could have limited generalizability}

% \subsubsection{Proactively helping users author constraints}

% \temp{things get long, then might forget}

% \temp{infer constraints from prompt}

%% file: sections/sec-060-conclusion.tex
\section{Conclusion}

In this work, we introduced a user-centered taxonomy of real-world scenarios, benefits, and preferred methods for applying constraints on LLM outputs, offering both a theoretical framework and practical insights into user requirements and preferences. In addition, we presented \systemname, an early GUI-based tool that enables users to prototype and test output constraints iteratively. Our results shed light on the future of more controllable, customizable, and user-friendly interfaces for human-LLM interactions.

%% file: sections/apx-methodology.tex
\section{The Survey Instrument}\label{sec:apx-survey-instrument}

In this section, we detail the design of our survey. The survey starts with questions about background and self-reported technical proficiency:

\begin{itemize}%[leftmargin=.2in]
    \item What best describes your job role: 
    software Engineer; 
    research scientist; 
    UX designer; 
    UX researcher; 
    product manager; 
    technical writer; 
    other (open-ended)
    
    \item To what extent have you designed LLM prompts: 
    a) I have ``chatted with'' chatbots like Bard / ChatGPT as a user; 
    b) I've tried making a prompt once or twice just to check it out, but haven't done much prompt design / engineering; 
    c) I have some experience doing prompt design / engineering on at least three LLM prompts;
    d) I have done extensive prompt design / engineering to accomplish desired functionality.
    Only those participants who selected either option c) or d) were given the opportunity to continue with the remainder of the survey.  This approach is specifically designed to exclude ``casual'' LLM users.
    
    \item I \emph{primarily} design prompts with the intent that they will be used by: 
    a) consumers / end-users (e.g. a recipe idea generator); 
    b) downstream development teams (e.g. captioning, classifiers); 
    c) both, I split my time about evenly between the two;
    d) other audience or use cases (open response).
\end{itemize}

The survey then asked participants to report \emph{three real-world use cases} where they would like to constrain LLM outputs. For each use case, participants were asked:

\begin{itemize}%[leftmargin=.2in]
    \item How would like to be able to constrain the model output (open response); 
    
    \item Provide a concrete example where it would be useful to have this constraint (open response);
    
    \item How precisely do you need this constraint to be followed: a) exact match; b) approximate match and why (optional open response);
    
    \item How important is this constraint to your workflow (5-point Likert scale from ``it's a nice to have, but my current workarounds are fine'' to ``it's essential to my workflow'') and why (optional open response).
\end{itemize}

The survey then asked participants to reflect through open response on scenarios where they would prefer \emph{expressing constraints via GUI} (sliders, buttons, etc.) over natural language (in prompts, etc.) and vice versa, as well as any alternative ways they would prefer to express constraints. 
To facilitate the reflection, the survey additionally asked participants to rate their level of preference in: 

\begin{itemize}%[leftmargin=.2in]
    \item Output should be exactly 3 words, no more than 3 paragraphs, etc. 
    \item Output in a specific format or structure (e.g., JSON, XML, bulleted / ordered list) 
    
    \item Only output ``left-handed'', ``right-handed'', or ``ambidextrous'' 
    
    \item Output must include or avoid certain words / phrases 
    
    \item Output must cover or avoid certain topics, only use certain libraries when generating code, etc. 
    
    \item Output style should mimic Yoda / Shakespeare / certain personas, etc. 

\end{itemize}

Each question presented a 7-point Likert scale from ``strongly prefer natural language'' to ``strongly prefer GUI.''